\documentclass[conference]{IEEEtran}
\IEEEoverridecommandlockouts
\usepackage{cite}
\usepackage{amsmath,amssymb,amsfonts}
\usepackage{algorithmic}
\usepackage{graphicx}
\usepackage{textcomp}
\usepackage{xcolor}
\usepackage{flushend}
\def\BibTeX{{\rm B\kern-.05em{\sc i\kern-.025em b}\kern-.08em
    T\kern-.1667em\lower.7ex\hbox{E}\kern-.125emX}}
\usepackage[utf8]{inputenc}
\usepackage{subcaption}
\usepackage{enumerate}
\usepackage{wrapfig}

\usepackage{xcolor,colortbl}
\definecolor{Gray}{gray}{0.85}
\begin{document}

\title{Experience with PCIe streaming on FPGA for high throughput ML inferencing}


\author{\IEEEauthorblockN{Piyush Manavar, Manoj Nambiar, Nupur Sumeet, Rekha Singhal, Sharod Choudhary, Amey Pandit}
\IEEEauthorblockA{\textit{TCS Research, Tata Consultancy Services, Mumbai, INDIA} \\
piyush.manavar[at]tcs.com, m.nambiar[at]tcs.com, nupur.sumeet[at]tcs.com, rekha.singhal[at]tcs.com,\\ sharod.rchoudhury[at]tcs.com, amey.pandit[at]tcs.com}

 }


\maketitle
\begin{abstract}
Achieving maximum possible rate of inferencing with minimum hardware resources plays a major role in reducing enterprise operational costs. In this paper we explore use of PCIe streaming on FPGA based platforms to achieve high throughput. PCIe streaming is a unique capability available on FPGA that eliminates the need for memory copy overheads. We have presented our results for inferences on a gradient boosted trees model, for online retail recommendations. We compare the results achieved with the popular library implementations on GPU and the CPU platforms and observe that the PCIe streaming enabled FPGA implementation achieves the best overall measured performance. We also measure power consumption across all platforms and find that the PCIe streaming on FPGA platform achieves the 25x and 12x better energy efficiency than an implementation on CPU and GPU platforms, respectively. We discuss the conditions that need to be met, in order to achieve this kind of acceleration on the FPGA. Further, we analyze the run time statistics on GPU and FPGA and identify opportunities to enhance performance on both the platforms. 
\end{abstract}

\begin{IEEEkeywords}
High performance computing, Field programmable gate arrays, Machine learning, 
\end{IEEEkeywords}

\section{{Introduction}}

ML-based inferencing is increasingly being used in enterprise systems being deployed in the cloud and on-prem data centers for real-time decision-making. Take the case of an online application that caters to a large number of concurrent users which could be in the range of 10K to 100K. If any kind of recommendations needs to be made to these users, real-time inferencing based on compute-intensive machine learning models could find scaling on general purpose CPU-based servers an expensive proposition. In retail scenarios for user-product ML models, inferences would need to be invoked many times per user. In such cases, machine learning models account for user characteristics as well as product characteristics then for each user the inferencing will need to be invoked as many times as there are products. This increases compute demands manyfold. For such systems, it is essential to have the capacity to serve with the maximum possible inferencing rate. This in turn helps reduce the cost of hardware to serve peak workloads.

In this paper we will evaluate these metrics on the following hardware platforms CPU, GPU and FPGA. The GPU and FPGA hardware evaluated are available as PCIe cards which can be fitted on the PCIe slots of the CPU based server. We find that PCIe streaming capability on FPGAs can influence high throughput rates and will discuss conditions that need to be met to leverage it. We also find that our FPGA implementation achieves very high throughputs at very low batch sizes, making it more suitable for interactive and bursty workloads.

The machine learning model used for the study will be an in-house gradient boosted decision tree model trained using the default configuration of 100 trees each with a maximum depth of 3. This model is trained with the PAKDD 2017 data set~\cite{Recobell}. The accuracy of this model is reported in~\cite{iPrescribe_paper}. While our implementations are verified functionally to give the same level of accuracy, any further discussion on model accuracy is out of scope of the paper.

The paper is organized as follows. Sections 2 and 3 cover training of the model, the relation of the data sets to retail workload scenarios and CPU implementation. Sections 4 describes hardware architecture of XGBoost. The FPGA implementation and it's integration are covered in section 5 \& 6. In section 7 GPU implementation is explained. Section 8 describes the experimental setup for performance and power. Section 9 analyses the results and section 10 makes projections of performance based on various bottleneck conditions.Section 11 discusses related work and section 12 concludes the paper.


\begin{figure*}
    \begin{minipage}{0.35\linewidth}
        \includegraphics[width=1\linewidth, height = 2.8cm]{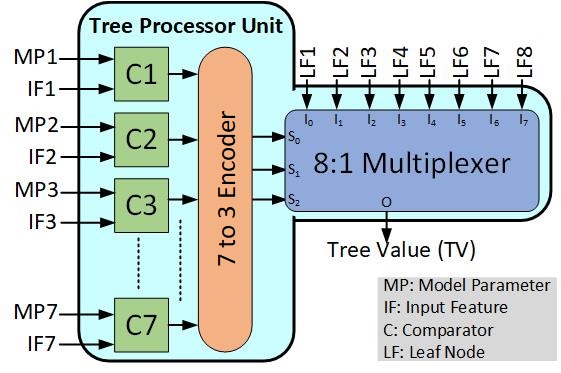}
        \caption{Digital circuit implementation of a Tree}
        \label{Fig1}
        \vspace{0.4cm}
        \includegraphics[width=1\linewidth]{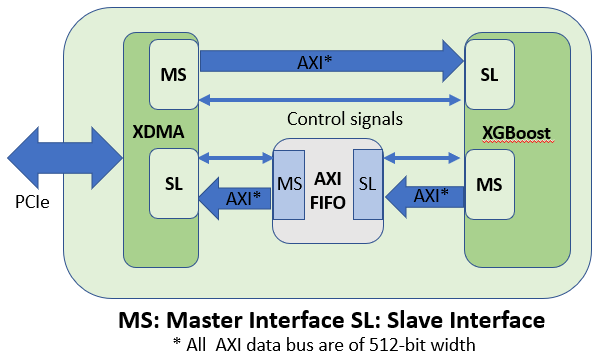}
		\caption{Implementation of XGBoost with PCIe.} 
		\label{Fig4b}	
    \end{minipage}
    \hspace{0.8cm}
    \begin{minipage}{0.65\linewidth}
        \includegraphics[width=1\linewidth, height = 6cm]{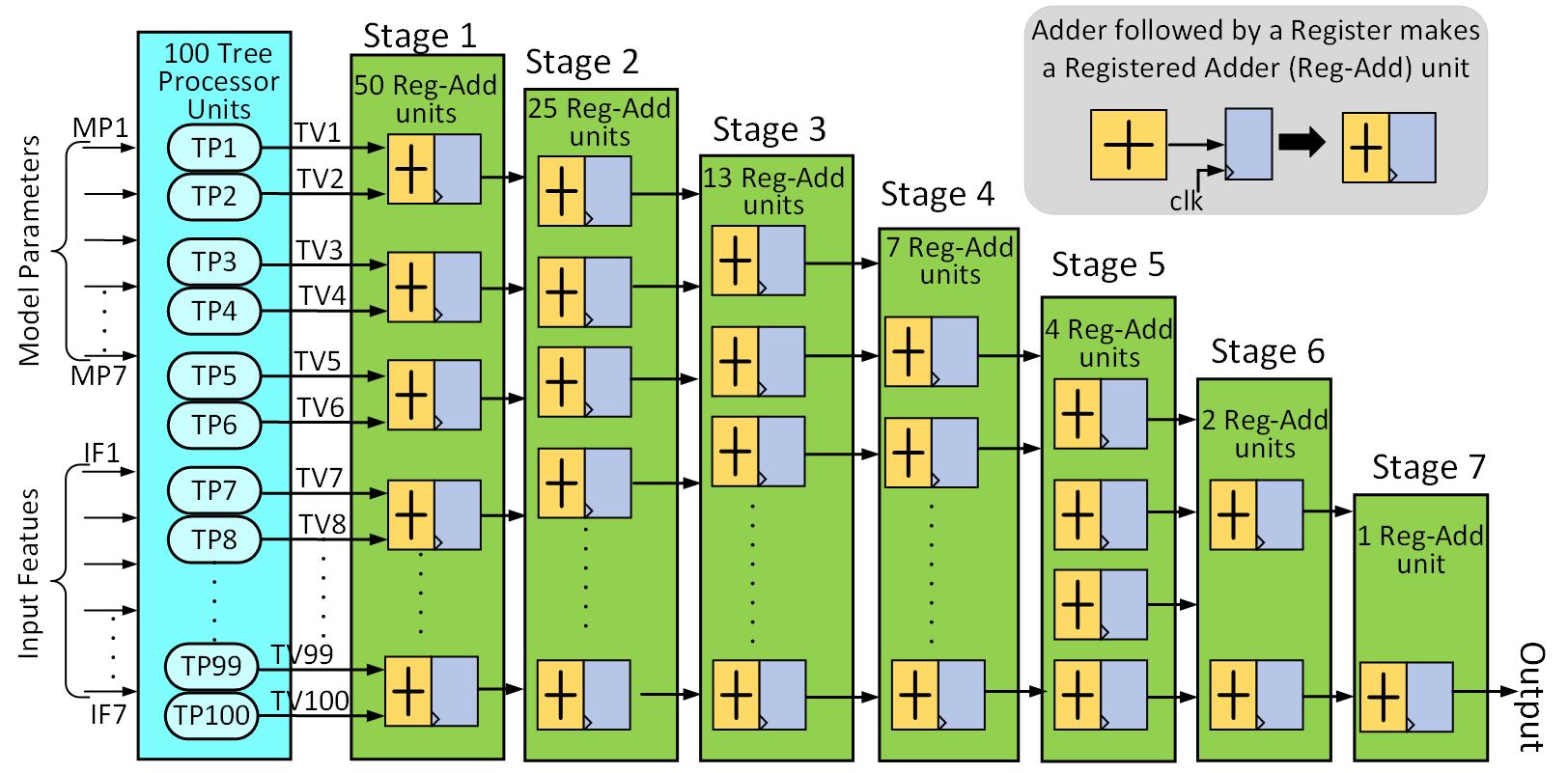}
        \caption{RTL realization of XGBoost}
        \label{Fig3}
    \end{minipage}
\end{figure*}
 

\section{{XGBoost Model Building}}
In order to prepare the training data, we use the iPrescribe framework~\cite{iPrescribe_paper}. This framework joins all the data provided by the Recobell~\cite{Recobell} data set and converts it into a common data format. It comprises of the attributes available in the original data set, of which the number of such unique attributes are $\approx$20. The feature engineering applied on this data, generates a set of additional (derived) features from the data. 

At the end of the feature engineering step there were a total of 1146 features. For each training record, the ground truth was the binary variable which indicates if the user purchased the product. 

This processed data with 1146 features was passed to the XGBoost machine learning algorithm. There were a total of 280,000 records. The data was divided into train and test data wherein 90\% of the data was used for training and the remaining 10\% for testing. The training concluded with an accuracy of 0.71 (AUC) on the test data set. It is further found that only 112 of the features were relevant.

The resulting XGBoost model has about 100 decision tree with a maximum depth of 3 and takes one or more features as input. Traversing the tree from root node to the leaf node depends on the value of Input Features (IF) and Model Parameters (MP). The input features are compared with the relevant node parameter. The result of the comparison is binary and decides the direction of tree traversal. Once a leaf node is reached it determines a numerical (real) value, which we shall refer to as Tree Value (TV).
Since the trained XGBoost model has 100 trees, after the tree traversal we have 100 TVs. These TVs are added to produce the final output which indicates the probability of the user making the order.

\section{{CPU Implementation}}
The CPU implementation of XGBoost was in Python. The default XGBoost library was used. Though program was called in Python, the XGBoost inference algorithm is itself implemented in C on a 48-core server. Using 48 parallel processes which are running iteratively , inference of a fixed batch size enabled by the multiprocessing library in Python we were able to measure upto 450,000 inferences per second overall with all CPUs at full utilization. This would amount to serving 9 users per second if an inference has to be made in a retail environment represented by the Kaggle dataset\cite{kaggle} with 50,000 products. Supporting more users during a busy hour would mean more servers and thereby resulting in high cost.

This motivates exploring the use of FPGAs and GPUs in recommendation systems which provides large computing resources at a comparatively affordable cost. For evaluating XGBoost inferencing on GPU, we choose to use the popular Forest inference library\cite{rapid_forest_ml_github}. For FPGAs, we developed a custom RTL (Register Transfer Logic) based digital architecture and is discussed in the forthcoming sections.


\section{{XGBoost: Hardware Architecture}}

FPGAs have the capability to support fine-grained parallelism. The 100 trees in the XGBoost model are independent of each other and this structure is amenable to parallelism. For the target XGBoost model, we developed an RTL design that would exploit maximum parallelism and execute the algorithm with minimum latency and consequently yield maximum throughput.

 Every decision tree can be represented as a table which is indexed by a combination of results of input feature comparisons with the tree node parameters. The maximum number of comparisons in a tree of depth 3 is 7. All comparisons can be scheduled in parallel since they are independent of each other. Each comparison operation is mapped to a 2-input comparator whose output is a binary value. A combination of these values could be encoded using a tree specific encoder to index one of the 8 possible leaves of the tree. The leaf values of the tree are fed as input to an 8:1 multiplexer. The select input of the multiplexers come from the encoder. Figure \ref{Fig1} shows the Tree Processing Unit that comprises of a set of comparators feeding a multiplexer through an encoder. The output of the multiplexer represents the tree value (TV). The model parameters values and input features are stored in registers. The XGBoost model contains 100 trees, each mapped to a Tree Processing Unit (see Figure \ref{Fig3}) and generate 100 TVs which are passed to an adder unit. We use a 7-stage adder to accumulate the tree values and generate the final result. Furthermore, we introduce pipelining in the adder unit by introducing registers after every adder stage. Because of this, the design can process one inference every clock cycle and primarily helps in improving the throughput of the architecture.

\section{{ FPGA Implementation}}
For data center use, FPGAs are available as Peripheral Component Interconnect Express (PCIe) based accelerator cards. The main program which is accelerated will always be invoked from the host CPU. It marshals the input parameters and passes them to the FPGA via the PCIe interface. The digital circuit implemented on the FPGA will process the inputs and pass the results back to the CPU via the PCIe interface. The device driver provided by the FPGA vendor is used to communicate with the FPGA accelerator card.

While we have described the XGBoost digital circuit in Figure \ref{Fig3}, the implementation is not complete unless there is the capability to interface with the PCIe bus. This is enabled by the use of the Xilinx provided XDMA IP~\cite{XDMA} block which can interface the application with PCIe. This IP helps to interface the application module (XGBoost) to the PCIe bus for communication with the software running on the host CPU. There are two ways in which the XDMA interface can be configured - memory mapped mode and streaming mode. In the memory mapped mode, the following set of operations execute serially for a single batch of inputs.
\begin{enumerate}
    \item   Copy from Host to FPGA memory
    \item  Processing of inputs by the XGBoost module
    \item  Copy of the inferencing results from FPGA memory back to host memory
\end{enumerate}

\begin{figure}[htb!]
	\centering
	
		\begin{subfigure}[b]{0.49\textwidth}
			\centering
			\includegraphics[width=\textwidth]{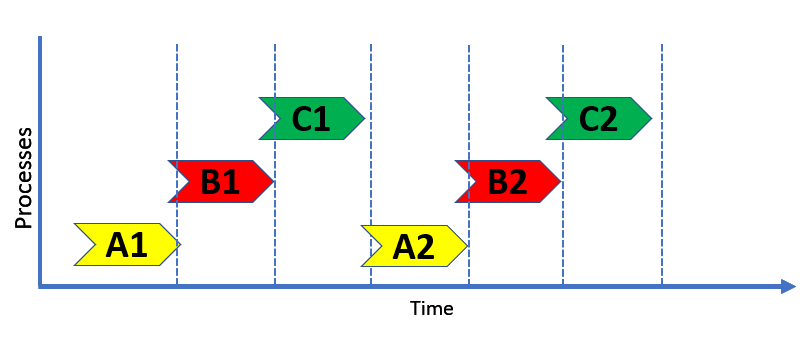}
			\caption[]%
			{{\small Serial execution in memory mapped configuration}}    
			\label{serial_mm}
		\end{subfigure}
	
		\vspace{0.5cm}
		\begin{subfigure}[b]{0.49\textwidth}  
			\centering 
			\includegraphics[width=\textwidth]{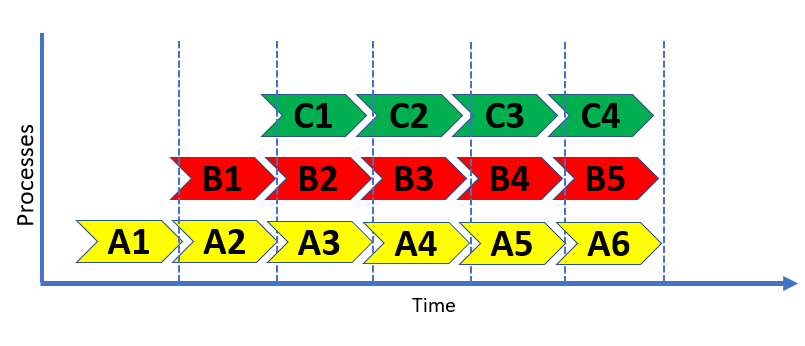}
			\caption[]%
			{{\small Pipe-lined execution achievable in memory mapped configuration}}    
			\label{pipeline_mm}
		\end{subfigure}
	
	\vspace{2mm}
	\begin{minipage}{0.8\linewidth}
		\emph{Ax: Data Transfer from Host to Device }
	\end{minipage}
	\begin{minipage}{0.8\linewidth}
		\emph{Bx: Data Processing}
	\end{minipage} 
	\begin{minipage}{0.8\linewidth}
		\emph{Cx: Data transfer from Device to Host}
	\end{minipage}
	
	\vspace{4mm}
	\caption{Memory mapped execution}
	\label{globallable}	
\end{figure}

 This can be visualized in Figure~\ref{serial_mm}. Multiple batches of inputs can be processed in a pipeline in such way that all the above tasks run concurrently, but on a different set of inputs, with one task feeding inputs to the next in the pipeline. This can be seen in Figure ~\ref{pipeline_mm}. This mode of memory mapped I/O is available in the GPU as well and the pipelined parallelism thus achievable using a feature called CUDA streaming~\cite{NVIDIA_Streams_and_Concurrency}. Please note that the maximum pipeline depth that can be achieved in this memory mapped mode of transfer is limited to 3, and every stage works on a batch of inputs. A careful tuning of batch sizes is essential to realize maximum inferencing speed.

\begin{figure}[htb!]
	\centering
		\includegraphics[width=1\linewidth, height = 6.5cm]{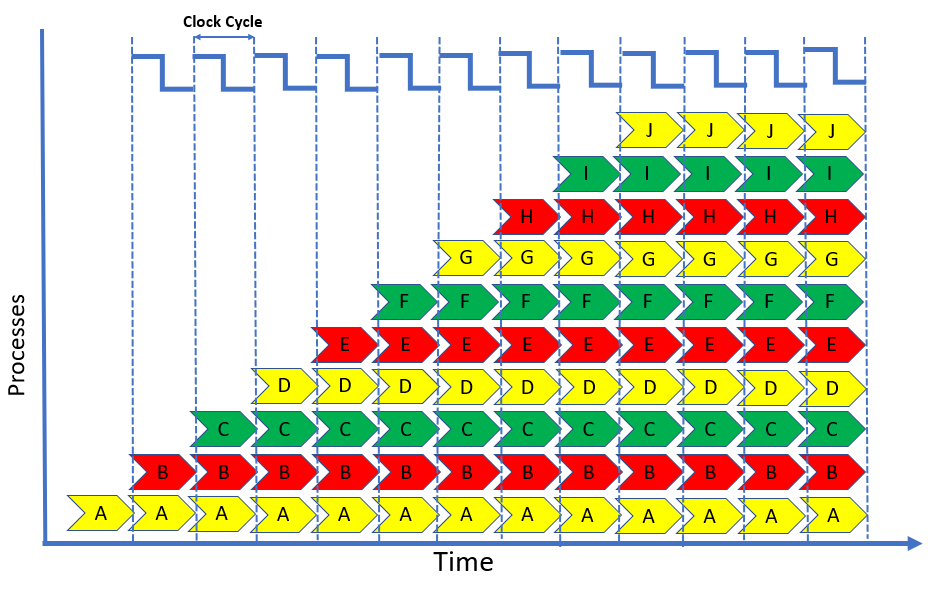}

        \begin{minipage}{0.8\linewidth}
        \emph{Ax: Data Transfer from Host to Device }
        \end{minipage}
        \begin{minipage}{0.8\linewidth}
        \emph{Bx: Tree processing}
        \end{minipage} 
        \begin{minipage}{0.8\linewidth}
        \emph{Cx - Ix: 7-stage parallel addition}
        \end{minipage}
		 \begin{minipage}{0.8\linewidth}
        \emph{Jx: Data transfer from Device to Host}
        \end{minipage}
        
        \vspace{2.5mm}
		\caption{Pipe-lined execution achievable in XDMA IP configured in streaming mode} 
		\label{Parallel_process_execution_in_FPGA}	
\end{figure}

\vspace{1mm}
 When the XDMA IP is configured in the streaming mode, there is no need to stored the data in FPGA memory. The data coming from the host CPU is passed directly to the application module (XGBoost in this case) and the output of the application module can be passed directly to the CPU via the XDMA interface. Theoretically it is possible for an input from the CPU to be presented to the application module every clock cycle, and same way in the reverse direction. In such a scenario, if the application is capable of processing a new input in every clock cycle, then it is possible to achieve a finer and higher level of pipelined parallelism. This is illustrated in the Figure~\ref{Parallel_process_execution_in_FPGA}, wherein the  system achieves a finer and deeper pipeline of 10 stages (8 stages of XGBoost + xdma input + xdma output)  (Figure~\ref{Fig3}). This level of fine grained pipelining makes the inferencing speed less dependent on input batch size. In this case, the theoretical maximum throughput achievable is equal to the frequency with which the application is synthesized on the FPGA. For e.g, if the application is synthesized at 250 MHz then the system bounded within the FPGA can deliver 250 million inferences per second. Given that the XDMA IP can deliver 64 bytes in a clock cycle a frequency of 250 MHz translates to 16 GB/sec which is equal to the PCIE v3 x16 speed in one direction. This leaves the onus on the CPU application and xdma device drivers running on the CPU to deliver the same speed to get the entire system (not just the FPGA) to deliver 250 million inferences/sec. 
 
 Additionally, we also configured the design with an additional AXI FIFO as shown in  Figure~\ref{Fig4b}. The FIFO acts as a buffer to store temporarily store model results in its way back to the CPU via the XDMA IP. It also makes interfacing with the XDMA IP simpler. This will add additional stages to the pipeline in figure 5, limited by the maximum depth of the FIFO, but would not affect the maximum achievable throughput.



 \vspace{2.0mm}

%
    

\vspace{1.0mm}
We developed an automation script that creates a HDL (Hardware Description Language) code for the XGBoost module. The 100 trees (with a depth of 3) of the XGBoost module are represented in the form of if-else branching statements in verilog.. The model parameters going as input to the tree nodes are ascertained from the re-trained (with 112 feature values) model weight file. This helps identify the useful model features and make efficient use of communication bandwidth between the FPGA and the host.

\section{{Integration with software}}
\begin{figure}[!ht]
	\begin{center}
	\includegraphics[width=0.8\linewidth, height = 4cm]{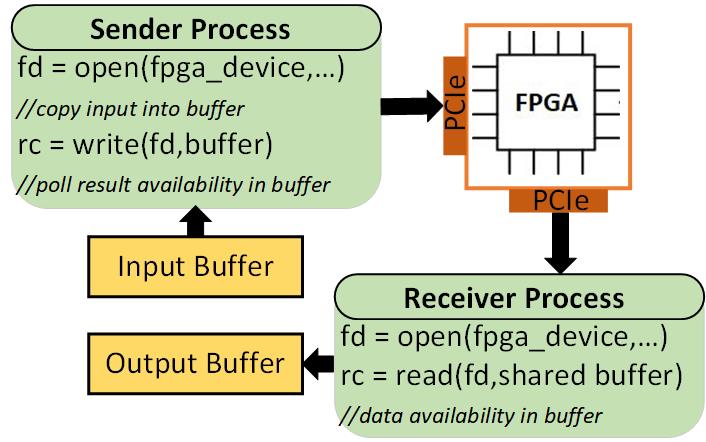}
\caption{Software interaction with FPGA} \label{Fig5}
	\end{center}
\end{figure}

Given the potential to transfer data and compute in pipe-lined parallelism, the software has to be able to exploit it. In order to achieve this we need two processes which can run in parallel, one to send the input data to the FPGA and another to collect the output results. The Sender process can be the application process which marshals the parameters and sends the data to the FPGA. The Receiver process could be a daemon process which reads the output results from the FPGA while copying it into a shared memory region that is visible to the application process. This is depicted in  Figure \ref{Fig5}
\vspace{3.0mm}

\vspace{3mm}
The Sender process in order to send the input data to the FPGA over PCIe invokes the write system call after opening the device descriptor. This is enabled by the device driver which is available from the FPGA vendor. On the other hand, the Receiver daemon process invokes the read system call to read the results coming from the FPGA.

\section{{GPU implementation}}
\label{section:my}
For studying, the XGBoost inferencing performance on a GPU, we used the Rapid  the Rapids Forest inference library~\cite{rapid_forest_ml_lib}. There is no PCIe streaming (like what was implemented using xdma streaming) on the GPU. So execution model of inference will be similar to what was discussed in the xdma memory mapped implementation, except that the GPU tree processing kernel will process the input. To calculate the throughput, we are noting  the timing of the predict function~\cite{rapid_forest_ml_lib} while varying the batch size. This includes memory transfers between the host CPU and the GPU.

However, the NVIDIA CUDA software framework allows for a different concept of streaming~\cite{NVIDIA_Streams_and_Concurrency} as discussed in section 4. Input data can be split into smaller batches. With streaming, it is possible that when a GPU kernel is processing input data from a previous stream, a data transfer from host CPU memory to GPU memory can be executed in parallel for the next stream. This kind of overlap can help achieve higher throughput. The Rapid Forests Inference Library does not explicitly support use of streams by the calling program in the python interface.
The Rapid Forests Inference library implements a number of optimizations to achieve a high speed GPU implementation of decision tree inferencing. The techniques that help increase the speed of inferencing on the GPU are well explained in~\cite{NVIDIA_Streams_and_Concurrency}.

\section{{Experiments}}
\subsection{Performance Measurement Setup}
The FPGA experiments were carried out on 2 socket HP DL380 Gen9 with 48 cores. The FPGA used was the Alveo U280 board which was inserted into a PCIe Gen3 $\times$16 slot in the server. The circuit as shown in Figure \ref{Fig4b} is synthesized and implemented on the FPGA at a frequency of 250 M\textit{Hz} using Vivado 2019 and PCIe IP DMA/Bridge Subsystem for PCI Express v4.1\cite{XDMA} and XDMA driver v2018. The FIFO maximum depth is set to 16. The resulting bit-stream is used to program the FPGA. Once the FPGA is programmed and the device drivers are loaded the FPGA subsystem is ready for operation. For taking the performance measurements, first the receiver process is started and immediately afterwards the sender process is forked. The sender process then sends a fixed set of input data to the FPGA. As soon as the receiver process receives all the expected results, it exits. The start of sender process and end time of receiver process are used to calculate the inference throughput.

Each feature was encoded to 4 bits in size in the FPGA implementation. Accordingly, an input with 112 feature vectors will require 448 bits or 56 bytes. The XDMA core was configured for a width of 512 bits (64 bytes). In our implementation, one input is transferred every clock cycle. So out of 64 bytes, 56 bytes would contain valid input data. 

In the course of testing, we encountered a problem with input batch size. We realized that the write and read calls (illustrated in Figure ~\ref{Fig5}) were not very dependable for input sizes larger than 1 MB (roughly 16000 inputs). So batch sizes larger than 15000 are simulated with multiple write and read calls by the relevant processes. The entire implementation results in very low utilization of the FPGA resources. The Combinational Logic Blocks (CLBs) were the most utilized resource for our design with a utilization slightly less than 7\%.

For the GPU measurements we used the DGX-1 server~\cite{DGX-1_V100}. This server had 8 Nvidia V100 GPUs with PCIe Gen3 $\times$16 connections. The DGX-1 server was running the Ubuntu operating system. For taking performance measurements, we used a docker image which had Rapid Forests inference library 0.19.0 and CUDA 10.2 installed. A Python program invoked from the docker instance is used to load the pre-trained XGBoost model file and the test inputs were also read from a file. The given batch of inputs from the file were used to invoke the Rapid forest XGBoost implementations. The time taken for the inference for the batch of inputs were recorded to calculate the throughput. The start and end timestamps are wrapped around the predict function of the forest inference library~\cite{rapid_api}

All the throughput measurements presented in this paper is an average taken over 10 measurements. Any deviations from this will be highlighted.

\subsection{{Power Measurements}}
For power measurement on the CPU and FPGA platforms, we use the HP iLO 4~\cite{ILO_4} Power meter. As fine grained power readings were not possible, we iteratively ran the workload for atleast 5 mins, before taking the readings. The power measured indicates the total server power consumption, including that of the FPGA accelerator card. The CPU and the FPGA experiments are run on different servers which have the same configuration. The idle power of server with FPGA was 135 W and the one w/o FPGA is 133 W. Using this information one could infer that the idle power consumption of the FPGA card is 2W. everything else being the same. We were using the DGX system for our GPU experiments and used the nvidia-smi~\cite{nvidea_smi} command to log power draw every 10 ms. Here again we repeat the test to last for more than 2 minutes. We calculate the average power during the test run to report power consumed during the experiment. Please note that this only has the GPU power consumption unlike that of the FPGA which includes the power consumption of 2 CPUs (sender and receiver process). The idle power consumption on the GPU was 43W.

\section{{Results and Analysis}}

Table ~\ref{through_put_table} displays the throughput achieved by the CPU, GPU and FPGA implementations as the input batch size is varied for the trained XGBoost model.

\vspace{7.0mm}
\begin{table}[htb!]
\centering
\caption{\label{through_put_table}  Inference measured for FPGA, CPU and GPU}
 \begin{tabular}{l c c c} 
 \hline
\rowcolor{Gray}\textbf{Batch Size} & \textbf{FPGA Inf*} & \textbf{CPU Inf*} & \textbf{GPU Inf*} \\
 \hline\hline
 1 & 2.2e-3 & 8.6e-3 & 0.001 \\ 

 10 & 0.02& 0.07 & 0.013 \\

 100 & 0.19& 0.72& 0.122\\

 1000 & 1.75& 2.78 & 0.810\\

 10000 & 6.55& 4.70 & 1.497\\ 

 100000 & 65.80& 4.85 & 1.425\\
 
 1000000 & ** & ** & 5.161 \\ [1ex] 
 \hline
 \multicolumn{4}{l}{* in million inferences/s}\\
 \multicolumn{4}{l}{** Results are not available because of system is not able to run at } \\
 \multicolumn{4}{l}{that batch size.}
\end{tabular}
\end{table}

\textbf{Single inference:} For a single inference (batch size 1) is measured to be 460 $\mu$s which includes the round trip latency of PCIe communication and FPGA processing.

The latency for XGBoost inference in the FPGA is only 9 clock cycles as can be observed from Figure ~\ref{Fig3}. This translates to 36 ns at 250 M\textit{Hz} which is negligible compared to the 400+ $\mu$s overall latency reported which includes communication over PCIe.

 The latency for a single inference in the GPU implementation is measured as 833 $\mu$s, nearly twice as that observed for a single inference on the FPGA. In the CPU we measured a single inference latency of 7 ms, which is very high latency on CPU as compared to the FPGA and the GPU. This exposes the high overhead in the CPU implementation.

\textbf{Batch Inference:} On all platforms CPU, GPU and FPGA we can see that batching inputs helps amortize overheads observed in single tests. We can see that the FPGA implementation achieves a maximum throughput of 65 million inferences/sec for a batch of 100k. This makes best use of the pipe-lining provided by the FPGA hardware design with the XDMA IP configured in streaming mode. This is slightly less than 1/3 rd of the maximum possible throughput permitted by a PCIe Gen3 $\times$16 connection. 

When we compare the FPGA and the GPU performance, we can see that the FPGA implementation achieved a throughput of 6.5 million inferences per second, for an input batch size of 10k, while the GPU implementation achieves a maximum throughput of 5.1 million inferences per second for an input batch size of 1 million. While the performance numbers are comparable, it is very important to note that the FPGA implementation achieves higher throughput at an input batch size which is two orders of magnitude less than the input batch size in which the GPU Rapid Forest library implementation achieves highest throughput. This makes the FPGA implementation more suitable for serving real time recommendations to online retail users. 

\textbf{Power Measurement:} We measured an average power consumption of 370W in the server when running the CPU experiment. On the FPGA server we measured an average power consumption of 195W. It is clear that FPGA scores better than the CPU in terms of power consumption as well. Considering that there are 48 CPUs each CPU core contributes roughly 5W power. Given that there are two processes running at 100\% CPU utilization we can infer that approximate power consumption by the FPGA to be 50W. Using nvidia-smi on the GPU being used we measured a power consumption of 58W. In order to compare this with the CPU only and FPGA experiments we will assume the the GPU is installed in a similar server which will consume an idle power of 133 W by the CPUs. Given that one CPU process is active during the benchmark we add 5W of power for the 1 CPU. Along with the 58W GPU power consumption the overall power consumption of a server with GPU can be estimated to be 197 W. When correlated with the performance results we can see that the FPGA has the best performance 337 K inferences per watt followed by GPU and CPU implementations in that order, with 26 K inferences per watt and 13 K inferences per watt respectively. This is summarized in Table ~\ref{Power_table}. We thus conclude that the FPGA platform achieves the 25X and 12X better energy efficiency than the CPU and GPU platforms, respectively. 
\vspace{2.8mm}
\begin{table}[htb]
\centering
\caption{\label{Power_table} Power cousumption for FPGA, CPU and GPU}
\vspace{1.50mm}
\begin{tabular}{c c c} 
\hline
\rowcolor{Gray}\textbf{ } & \textbf{Power (Watt)} & \textbf{Performance (inf/Watt)}\\
\hline\hline
CPU & 370 W & 13000 inf/W\\
FPGA & 195 W & 337000 inf/W\\
GPU & 58 W & 26000 inf/W\\
\hline
\end{tabular}
\end{table}

\section{{Performance Analysis}}
\textbf{GPU Performance Analysis:} The 3 major operations being performed during GPU inference are:
\begin{enumerate}
	\item copy of data from host memory to GPU memory (H2C) 
    \item launching of the tree processing kernels (K)
    \item copying of results from GPU memory back to CPU memory (C2H).
\end{enumerate}
For a batch of one million inferences (last row in table 1), as observed using the nvprof  tool, these tasks consume 40 ms, 2.84 ms and 3 ms, respectively. The profile revealed that these operations were not pipelined. In other words the execution followed the same profile as Figure~\ref{serial_mm}. The overall time consumed for the processing of the entire batch is 193 ms. This shows that the CPU processing contributes as the major part of the bottleneck. This includes the conversion of the input record in pandas [25] format being converted to the one which could be processed by the GPU kernel. Another interesting observation from the nvprof tool is the bandwidth achieved during the C2H operation which is reported as 10.33 GB/sec, with the size of the transfer being reported as 427.5 MB. Given a batch of 1 million inputs this translates to an average of roughly 420 bytes per input record. Given 112 features per input it seemed to be taking nearly 4 bytes to represent each feature. In the FPGA design only 4 bits per feature were used bringing the size of the input to 56 bytes\footnote{This was rounded up to 64 bytes to keep it compatible with the 512 bit interface of the XDMA IP which helped keep the FPGA design simple.}.

\textbf{FPGA performance analysis:} The inference on FPGA utilized the streaming PCIe interface. A maximum throughput of 65 million messages per second means that the speed achieved using PCIe in either direction is a little more than 4 GB/s with 64 bytes used per input in the current set up. To check if this was a limitation of our inferencing implementation, we tested with a loop back application, in which the XDMA input and output interfaces are directly connected to each other. This can be visualized in Figure \ref{Fig4b}, minus the XGBoost and AXI FIFO modules, with the XDMA master interface directly connected to the slave interface. This application was available as an example design from Xilinx \cite{XDMA}. In this test, we were able to achieve a maximum overall throughput of 5.5 GB/sec. Given that the PCIe interface could theoretically support a maximum speed of 15 GB/sec this was only 1/3$^{rd}$ of the maximum capacity. Thus, we observe, a 30\% performance overhead introduced by the XGBoost design compared to the plain loop back test. Even in the loop back test only 1/3 rd of the available PCIe bandwidth was achieved. As both the designs were synthesized at 250 MHz they were both capable of maxing out the PCIe bandwidth at 15 GB/sec. These observations could be the result of complex system interplay including the device drivers, the XDMA IP,the XGBoost hardware block and the AXI FIFO. This needs to be further investigated.

  From the measurements we can see that the FPGA based implementation outperforms the GPU implementation. However the following factors need to be kept in mind
\begin{itemize}
    \item The processing of 100 trees is being done in parallel in both FPGA and GPU. Given this high speed processing the bottleneck lies in the PCIe communication with the CPU in both cases.
    \item The FPGA implementation achieved a pipelined parallelism with 3 tasks running in parallel - the sender process (CPU), the XGBoost computation (FPGA) and the receiver process. However in the case of the GPU implementation using the Rapid Forests Library there was no such parallelism - at any point in time it would be either the CPU or the GPU or the data transfer between the two. 
    \item The PCIe based streaming capability enabled by the XDMA IP made a significant impact to the measurements.
    \item The GPU based system was able to utilize higher PCIe bandwidth of 10.3 GB/sec, while the FPGA based implementation could achieved only 4 GB/sec, with both systems having the same PCIe capacity PCIe v3 x16
    \item The input record size in the GPU implementation was 420 bytes , while in the FPGA implementation it was 64 bytes
    \item The CPU processing proved to be the major overhead in the GPU implementation while in the FPGA implementation there was negligible CPU processing
    \item The FPGA implementation was customized to the workload at hand, whereas the GPU implementation used a popular library which could be used to implement any decision tree model.
\end{itemize}

While we can appreciate the impact of streaming on the performance of the FPGA implementation, we have to note that there is scope for additional performance engineering to speed up the inference rate on both the GPU as well as FPGA.

\textbf{Conditions for leveraging PCIe streaming for high performance:} It is important to note that PCIe based streaming may not be very useful for all workloads. In cases where there is very little scope for pipe lining in the hardware design. In such cases it will not be possible for sending records with a high rate without a queue building up within the FPGA. The FPGA implementation of the XGBoost model had an initiation interval (or II)~\cite{Initiation_Interval} of one. The implementation had the ability to ingest an input record every clock cycle and output a result every clock cycle without queue any queue buildup. A design with a very high II cannot benefit from PCIe streaming. This means that there is one bottleneck component in the design which takes two clock cycles to process every input. This will required that the data flow be unrolled as much as possible and fit within the FPGA chip estate. Hence smaller machine learning models are better poised to benefit from PCIe streaming on FPGA to contribute towards higher end to end throughput of applications. 

The energy efficiency observed on FPGA platform make them the most sustainable implementation for use in data centers for such models.

\section{{Related Work}}
Owaida et.~\cite{Owida_1} implement a route scoring application on the FPGA using decision tree ensembles like XGBoost. In~\cite{Owida_2} the authors discuss ways to partition large decision tree ensembles across multiple FPGAs (cluster). The work is extended to~\cite{Owida_3} wherein measurements are presented in various hardware configurations and various tree depths. These are based on the design of a custom decision tree processor, multiple instances of which are implemented on a single FPGA. The decision tree processor processes a tree, one at a time and implements pipe lined parallelism across trees.  In~\cite{rapid_forest_ml_lib} the authors describe the implementation of the decision tree inference on the GPU platform. They report results with XGBoost models trained using various data sets and their trained models are different in configuration to the one referred to in the paper. Moreover there is very little detail about what is actually being measured, because of which an apples to apples comparison is not possible. Hence we have refrained from extrapolation their results.

Authors in ~\cite{DrGautam_Rekha} discuss the methodology of recommending products to the retail end user customers, introducing a way of how past transactional data is processed, features engineered and used to train machine learning models which could be used to serve real time recommendations. Paper~\cite{iPrescribe_paper} describes an end to end implementation of such a system by leveraging the Recobell and the Kaggle Instacart data sets. They present how use of in memory technologies help build a scalable system on CPU based server clusters, with XGBoost (with the same configuration as evaluated in the paper) as one of the inferencing algorithms.

~\cite{Al_Enabling_Real_Time_Adaptivity_in_MOOCs} Describes the architecture and various components involved in making personalized recommendations to users in selecting the next course they would be interested in.

~\cite{R2SIGTP: a Novel Real-Time Recommendation System} Describes the same for an application which makes recommendations which accounts for the users geographical location. Both systems are similar to the application which we described in the sense that they make contextualized decisions in real time. However there is very little discussion on performance and neither papers share experimental performance evaluation results.

~\cite{hls4ml} implements decision tree using Boolean function and calculating the index of leaf. That index is used to address a look-up table to output the final value of leaf.

The unique contributions of this paper are as follows
\begin{itemize}
    \item Performance and and Power Comparison of proposed FPGA and popular GPU and CPU implementation for the same model trained from a given public dataset with experimental results
    \item Analysis of a streaming DMA implementation on FPGA for the model
    \item Performance Analysis of popular GPU and CPU implementations
   
\end{itemize}
\section{{Conclusion}}
While FPGA's are making their way into data centers we can see that there are opportunities where they could serve as practical accelerators for ML inferencing. We have taken the case of a lightweight recommendation model which will be useful in serving busy hour workloads for online retail enterprises. Based on our measurements we have seen that the FPGA implementation scores well when it comes to performance and power consumption. Models that can be implemented on the FPGA resources with low II,  have better prospects of leveraging higher performance on the FPGA with PCIe streaming. Based on the analysis of the underlying hardware capabilities, we also see that there is more scope to engineer performance on the GPU and the FPGA to achieve higher inferencing speeds.


\end{document}